\documentclass[useAMS,usenatbib,referee]{mn2e}
\usepackage{epsfig}
\title[A pulsar outer gap model with trans-field structure]{A pulsar outer gap model with trans-field structure}
\author[J.~Takata, S.~Shibata \& K.~Hirotani]
  {J.~Takata,$^1$\thanks{takata@ksirius.kj.yamagata-u.ac.jp}
  S.~Shibata,$^2$ K.~Hirotani,$^3$ \\
  $^1$Graduate School of Science and Engineering, Yamagata University, Yamagata
990-8560, Japan\\
  $^2$Department of Physics, Yamagata University, Yamagata 990-8560, Japan \\
  $^3$Max-Planck-Institut fuer Kernphysik, Postfach 103980 D-69029 Heidelberg, Germany }
\begin{document}
\date{}

\pagerange{\pageref{firstpage}--\pageref{lastpage}} \pubyear{2004}

\maketitle
\label{firstpage}
\begin{abstract}

We investigate the electrodynamics of an outer gap in the 
meridional plane of the aligned-rotator. 
The charge depletion from the  Goldreich-Julian charge density  causes 
a large electric field along the magnetic field line. The electrons or 
the positrons are accelerated by the field-aligned electric field and radiate 
the $\gamma$-rays tangentially to the local magnetic field line.
 Some of such $\gamma$-rays collide with $X$-rays to materialize as the  
electron-positron pairs on  different field lines from the field line on 
which they were emitted.  As a result, the electric field structure is 
expected to change across the field lines. Including these trans-field effects,
 we solve the formation of the electric field self-consistently with 
the curvature radiation and the pair creation processes. The $\gamma$-ray
 emission  and the pair creation are treated by use of Monte Carlo technique.  
We demonstrate that the distribution of the electric field along the 
field lines is affected by both the gap geometry and the external currents  
coming into the gap through the boundaries. 
In the electrodynamical model, it has been known that 
the solution disappears if the current density carried by the 
electron-positron pairs produced in the gap exceeds a critical value. 
We show that the critical current density is significantly increased when 
the trans-field structure is taken into account.
We also find that the location of the inner boundary of the gap shifts toward 
the stellar surface from the conventional null surface as the current density  
increases. The reason for the shift is derived from the stability condition 
of the inner boundary.  We also argue that the ideal-MHD condition holds  
outside of the gap only when the low energy particles coexist with
 the high energy particles migrating from the gap. 
\end{abstract}
\begin{keywords}
radiation mechanism: non-thermal - method: analytical - pulsar: general - $\gamma$-rays: theory
\end{keywords}
\section{Introduction}
The rotation-powered pulsars have been known as  sources of pulsed radio 
emission. The\textit{Compton Gamma-Ray Observatory} (CGRO) had 
shown that the young pulsars are also strong $\gamma$-ray sources, and 
 had detected seven $\gamma$-ray pulsars (Thompson 1999). 
The CGRO revealed the  fact that $\gamma$-ray luminosity accounts for 
about 10\% of the spin down energy.  
Furthermore, the observations for the  $\gamma$-ray spectra and 
light curves tell us a lot about acceleration of particles in 
the pulsar magnetosphere. The energy peak in the spectra above 1GeV 
shows that the electrons or the positrons are accelerated above $10^{12}$eV. 
A pulsation in the observed light curves implies that the acceleration 
of the particles and the subsequent radiation take place within the light cylinder, 
the axial distance of which is given by $\varpi_{lc}=c/\Omega$, 
where $\Omega$ is the angular frequency, and $c$ the speed of light.  
Although these observations provide constraints on proposed models, 
the origin of the $\gamma$-ray emission is not conclusive up to now.  

The global model for the acceleration of the particles in the magnetosphere 
has been considered as follows. 
The pulsar is an electric dynamo of about $10^{16}$V. 
The resultant electromagnetic energy is carried  into the 
magnetosphere by a longitudinal current. Then, if there is a large  electric
 potential drop along the magnetic field, 
the current carriers (probably electrons and/or positrons) should be 
accelerated to high energies by the  field-aligned electric field, 
and should release the electromagnetic energy by emitting  $\gamma$-rays. 
If there were no electric potential drop along the magnetic field, the 
space charge in the magnetosphere should be 
the Goldreich-Julian charge density (Goldreich-Julian 1969, hereafter GJ),
\begin{equation}
\rho_{GJ}=-\left(\frac{\Omega}{2\pi c}\right)\mathbf{e}_z\cdot
\left[\mathbf{B}-\frac{1}{2}\mathbf{r}\times(\nabla\times\mathbf{B})\right],
\label{rhogj}
\end{equation}
where $\mathbf{e}_z$ is the unit vector along the rotation axis, and
$\mathbf{B}$ is the magnetic field. 
On the other hand, any charge depletions from the 
GJ value cause the field-aligned electric field $E_{||}$. 

According to the above point of view, the acceleration of particles and 
the subsequent $\gamma$-ray emission have been argued with the polar cap model 
(Sturrock 1971; Ruderman \& Sutherland 1975) and the outer gap model  (Cheng, 
Ho \& Ruderman 1986, here after CHR),  the acceleration regions of which are, 
respectively, 
 located near the stellar surface above the magnetic poles and in the outer 
magnetosphere around the null charge surface ($\rho_{GJ}=0$) above 
the last open field lines. Because the magnetic geometry of 
the polar caps explains well the observed radio 
polarization, it is widely accepted that the radio
 emission occurs above the polar caps.

The outer gap model has been successful in explaining the observed 
light curves. By solving the Poisson equation for a vacuum gap, 
CHR concluded that the outer gap extends between the null 
surface and the light cylinder. With this vacuum gap geometry, the double 
peaks in the light curves can be interpreted as an effect of aberration 
and time delay of the emitted photons 
(Romani \& Yadigaroglu 1995; Cheng, Ruderman \& Zhang 2000). Although 
the spectrum has also been calculated 
with the models, in which $\gamma$-ray radiation has been calculated with 
an assumed field-aligned electric field, the models are not satisfactory 
in the sense of electrodynamics. Dyks \& Rudak (2003) argued that 
the acceleration region extends to the stellar surface and the 
light cylinder to reproduce the observed outer-wing emission, 
and the off-pulse emission in the Crab pulsar, 
questioning the traditional \textit{vacuum} gap geometry.

A non-vacuum outer gap model was proposed by Hirotani \& Shibata (1999).  They 
have focused on the electrodynamics, in which screening of the electric 
field by the electron-positron pairs is taken into account. In the traditional 
outer gap models such as Romani \& Yadigaroglu (1995),  distribution of
 the electric field was not determined 
 in a self-consistent manner with the radiation and pair creation processes. 
Hirotani \& Shibata (1999) solved the field-aligned electric field 
self-consistently with the curvature radiation and the pair creation processes,
 although they worked in one-dimension along the last open field line 
(see also Hirotani 2003 for recent version of the one-dimensional model). 
They argued that if there is an  injection of the particles into the gap, 
the position of the gap changes  outwardly or inwardly 
relative to  the null surface (Hirotani \& Shibata 2001). 

In these one-dimensional electrodynamical models, the current carried
 by pairs produced in the gap was restricted to be about 10\% of
 the GJ value.  So, the model does not account for the observed $\gamma$-ray 
fluxes of some pulsars, although reproducing the $\gamma$-spectra well 
(Hirotani, Harding \& Shibata 2003; Takata, Shibata \& Hirotani 2004). 
In the traditional outer gap model, on the other hand, the  current densities
are not calculated, but assumed to be the GJ values. The limitations  
on the current density and the $\gamma$-ray luminosity in the 
electrodynamical model may be due to one-dimensionality. It is obvious that 
the electrodynamics is affected by the trans-field effect such that 
the accelerated particles causes new pairs to be created on different 
field lines from the residing field line as pointed by CHR. 
Furthermore,  calculations for the light curves and the phase-resolved spectra 
require departure from the one-dimensional model. On these ground, 
in this paper,  we extend the one-dimensional electrodynamical model into a 
two-dimensional one. 

The most important trans-field  effect is caused by curved field 
lines. The $\gamma$-rays, which are emitted tangentially to the 
local field lines, may covert into the pairs on the different field lines from 
the field line on which they are emitted. In the present paper, 
to avoid additional complications by the  three-dimension, 
we  study a two-dimensional model.   In the subsequent paper, we 
will extend the two-dimensional model to a three-dimensional one. 
 Our aim  is to see how the structure of the electric field,
 the current density, and the geometry of the gap are affected 
by the trans-field effect. 

By neglecting the magnetic 
component generated by the current, we assume the dipole magnetic field,  
and we consider  the electrodynamics in the meridional plane of an 
aligned-rotator. This treatment simplifies the problem mathematically. 
Although it has been pointed out 
that an aligned-rotator is not active, that is, there are no 
longitudinal currents circulating in the magnetosphere 
(Michel \& Li 1999), the present frame work of the aligned models will apply 
to  \textit{nearly} aligned-rotators with the longitudinal currents. 

We find a solution in which the particles produced in the gap carry 
 about 30\% of the GJ current, which is significantly larger than the value 
$(\sim 10\%)$ obtained in the one-dimensional model. We also show that  
inner boundary of the gap shifts toward the 
stellar surface as the current increases. 
 
In \S\ref{model}, we present the basic equations for the electrodynamics 
in the gap, and describe treatment for the pair creation process. 
In \S\ref{result}, we show the results, and then discuss firstly  
the electric structure of \textit{vacuum} and secondly \textit{non-vacuum}
 cases. We also discuss the position of the inner boundary of the gap. In the 
final section, we compare the results with the previous works. We also discuss 
the dynamics for outside of the gap. 

In the following sections, we make statements not only 
about the aligned-rotators but also about inclined-rotators if applicable.
For definite of sign of charge, we assume ``parallel rotator'', for which 
the polar-caps are negatively charged. 
 
\section{Two-dimensional outer gap model}
\label{model}
The  present scheme consists of three main parts as follows;  
\begin{enumerate}
\renewcommand{\theenumi}{(\arabic{enumi})}
\item Poisson equation (\S\S\ref{poeq}) is solved numerically with proper 
boundary conditions to give the electric field,
\item the continuity equation for the particles (\S\S\ref{coeq}) 
with pair creation term is solved on every field line to give the space 
charge density,
\item $\gamma$-ray field and pair creation rate are obtained by using 
 Monte Carlo technique (\S\S\ref{boundary}).
\end{enumerate}
These processes are iterated until all quantities are self-consistent 
(\S\S\ref{method}).

\subsection{Poisson equation}
\label{poeq}
If  stationary condition ($\partial_t+\Omega\partial_{\phi}=0$, where 
$t$ and $\phi$ are the time and the azimuth, respectively)
is satisfied, that is, if there is no time variation of any quantities as
 seen in co-rotating system, then an electric field,
\begin{equation}
\mathbf{E}=-(\Omega\mathbf{e}_{z}\times\mathbf{r})\times\mathbf{B}/c
-\nabla\Phi_{nco}.
\label{elect}
\end{equation}
is exerted at position $\mathbf{r}$ (Mestel 1999). The electric field can be
 expressed as the sum of the corotational 
part, $-(\Omega\mathbf{e}_{z}\times\mathbf{r})\times\mathbf{B}/c$, 
and the non-corotational part, $-\nabla{\Phi}_{nco}$.  

Goldreich \& Julian (1969) proposed  that the pulsar magnetosphere is filled 
by  plasmas, and that the corotational  condition
\begin{equation}
\mathbf{E}\cdot\mathbf{e}_{||}=-\mathbf{e}_{||}\cdot\nabla\Phi_{nco}=0
\label{idmhd}
\end{equation}
 is satisfied, where $\mathbf{e}_{||}=\mathbf{B}/|\mathbf{B}|$ is the 
unit vector along the field line.  If the star crust is a rigid 
rotating perfect conductor, and if the condition (\ref{idmhd}) holds all the 
way along the field lines connecting the star and the region considered, then 
the uniform value of $\Phi_{nco}$ over the stellar surface propagates into the 
magnetosphere. We set an arbitrary constant for $\Phi_{nco}$ on the surface to 
be zero in the following. The electric field (\ref{elect}) becomes 
 $\mathbf{E}=\mathbf{E}_{co}\equiv-(\Omega\mathbf{e}_{z}\times\mathbf{r})
\times\mathbf{B}/c$, which  leads directly the corotational
 charge density (\ref{rhogj}).

In  a charge depletion region, where the condition (\ref{idmhd}) 
does not hold, $E_{||}$-acceleration is indicated in the equation 
(\ref{elect}), $E_{||}=-\mathbf{e}_{||}\cdot\nabla\Phi_{nco}\neq0$. 
By substituting equation (\ref{elect}) in $\nabla\cdot\mathbf{E}=4\pi\rho$, 
we obtain the Poisson equation of the non-corotational potential $\Phi_{nco}$, 
\begin{equation}
\triangle\Phi_{nco}=-4\pi(\rho-\rho_{GJ}),
\label{poisson}
\end{equation}
where $\rho$ is the space charge density, and $\triangle$ is the  Laplacian. 
 
If the variation in the azimuthal direction is negligible, we may reduce 
(\ref{poisson}) to 
\begin{equation}
\triangle_{r,\theta}\Phi_{nco}=-4\pi(\rho-\rho_{GJ}),
\end{equation}
 where $\triangle_{r,\theta}$ represents ($r,\theta$)-parts of
 the Laplacian; $r$ and $\theta$ are the radius and the colatitude, 
respectively, in the spherical polar coordinates. This approximation is justified only if 
the gap dimension 
in the meridional plane is much smaller than that in the azimuthal direction. 
We must lose some effects with this simplification, but we retain the 
trans-field effect, on which we put most interest. 

We adopt an orthogonal curvilinear coordinate system based on the dipole field 
lines. The coordinates ($\chi,\zeta$) are defined by 
\begin{equation}
\frac{r\sin^{-2}\theta}{\varpi_{lc}}=\textrm{constant along a dipole 
field line}\equiv \chi,
\end{equation}
and   
\begin{eqnarray}
\frac{r\cos^{-1/2}\theta}{\varpi_{lc}}&=&\textrm{constant along a curved line}
\\ \nonumber
&& \textrm{perpendicular to the field lines}\equiv \zeta,
\end{eqnarray}
respectively, where  $r$ is the distance from the centre of the star, 
and $\theta$ is the colatitude angle with respect to the rotational axis.  
The line $\chi=1$ draws the last open field line, which is tangent to the 
light cylinder, and $\chi=\infty$ corresponds to the magnetic axis or 
$r=\infty$. The Laplacian becomes 
\begin{equation}
\triangle_{r,\theta}=\frac{3\cos^2\theta+1}{\sin^6\theta}\frac{\partial^2}
{\partial \chi^2}+\frac{4}{\chi\sin^6\theta}\frac{\partial}{\partial \chi}
+\frac{3\cos^2\theta+1}{4\cos^3\theta}\frac{\partial^2}{\partial \zeta^2}
+\frac{3(3\cos^2\theta+1)}{4\zeta\cos^3\theta}\frac{\partial}{\partial \zeta}.
\end{equation}
 The  non-corotational electric field $\mathbf{E}_{nco}$ is obtained by  
\begin{equation}
E_{||}=-\frac{\partial\Phi_{nco}}{\partial s_{||}}=
-\frac{\sqrt{3\cos^2\theta+1}}{2\varpi_{lc}\cos^{3/2}\theta}
\frac{\partial\Phi_{nco}}{\partial \zeta},
\end{equation}
\begin{equation}
E_{\perp}=-\frac{\partial\Phi_{nco}}{\partial s_{\perp}}=
-\frac{\sqrt{3\cos^2\theta+1}}{\varpi_{lc}\sin^3\theta}
\frac{\partial\Phi_{nco}}{\partial \chi},
\end{equation}
where $ds_{||}=2\cos^{3/2}\theta/\sqrt{3\cos^2\theta+1}d\zeta$ and 
$ds_{\perp}=\sin^3\theta/\sqrt{3\cos^2\theta+1}d\chi$ are the line elements 
along the field lines and the perpendicular curved lines, respectively. 
Since dipole field is curl-free, we use $\rho_{GJ}\sim-\Omega B_z/2\pi c$ 
as the GJ charge density.
\subsection{Continuity equations for particles}
\label{coeq}
Let us denote  velocity $\mathbf{v}$ as 
\begin{equation}
\mathbf{v}=v_{||}\mathbf{e}_{||}+\mathbf{v}_d,
\end{equation}
where $v_{d}$ is the drift velocity owing to  
the electric field in the gap. The drift velocity is separated into the 
corotational and the non-corotational parts, 
\begin{equation}
\mathbf{v}_d=\mathbf{v}_{co}+c\frac{\mathbf{E}_{nco}\times\mathbf{B}}{B^2},
\label{drifteq}
\end{equation}
where $\mathbf{v}_{co}=\Omega\mathbf{e}_{z}\times\mathbf{r}$.
In the later section (\S\ref{disc}), we shall see that the magnitude of  
the non-corotational drift is negligibly small in comparison with 
the corotational drift.  By using $\nabla\cdot(n\mathbf{v})=
\mathbf{B}\cdot\nabla(v_{||}n/B)+\Omega\partial n/\partial\phi$ and  
imposing the stationary condition ($\partial_t+\Omega\partial_{\phi}=0$), 
the continuity equations become 
\begin{equation}
 \mathbf{B}\cdot\nabla\left(\frac{v_{||}N_{\pm}}{B}\right)=\pm S(\mathbf{r}),
\label{contin}
\end{equation}
where $S(\mathbf{r}$) is the source term due to the pair creation, $N_+$
 and $N_-$ denote the number density of outwardly and inwardly moving 
particles, respectively. We suppose that the accelerated particles have 
almost the speed of light, $v^2\sim v_{||}^2+v_{co}^2\sim c^2$, such that  
the parallel component is assumed to be $v_{||}=\sqrt{c^2-(\varpi\Omega)^2}$. 
With this parallel motion, the continuity equations (\ref{contin}) yield 
the particle's number density, provided that the source term 
$S(\mathbf{r})$ is given. 

If one assumes that the particle's motion immediately
 saturates in  balance between the electric and the radiation 
reaction forces, the Lorentz factor of the particles within the gap is 
simply given by 
\begin{equation}
\Gamma_{sat}(\mathbf{r})=\left(\frac{3R_c^2}{2e}E_{||}+1\right)^{1/4},
\label{lorent}
\end{equation}
where $R_c$ is the curvature radius of the magnetic field line. This
saturation simplifies the problem significantly. However, the particles in
the gap will have the saturated Lorentz factor $\Gamma_{sat}$ only when
 typical accelerating time $t_{ac}$, radiation damping time $t_{d}$ and
 crossing time $t_{cr}=W_{||}/c$, where $W_{||}$ is the gap width along the
magnetic field lines, satisfy the condition,
\begin{equation}
 t_{cr}\gg t_{ac}\ \ \textrm{and}\ \ t_{cr}\gg t_{d}.
\end{equation}
Hirotani, Harding \& Shibata (2003), who solved  unsaturated motion of
the particles, showed that the condition $t_{cr}\gg t_{ac}$ is
not satisfied effectively for some pulsars. Although
the unsaturated motion affects the $\gamma$-ray spectrum below 1GeV
(Hirotani et al. 2003; Takata et al. 2004), effect on the electric field
is less important. As far as the electrodynamics are concerned,  the 
assumption of the saturation will provide a good approximation.

The source term can be expressed as
\begin{equation}
S(\mathbf{r})=\int_0^{\infty}d\epsilon_{\gamma}[\eta_{p+}G_{+}+\eta_{p-}
G_-],
\end{equation}
where $\epsilon_{\gamma}$ is the photon energy normalized by $m_ec^2$,
$G_{+}$ and $G_{-}$ denote the distribution function of outwardly and inwardly
propagating $\gamma$-rays. The pair creation rate $\eta_{p}$   per unit
time per $\gamma$-ray photon with the energy of $\epsilon_{\gamma}$ is given by
\begin{equation}
\eta_p=(1-\mu_c)c\int_{\epsilon_{th}}^{\infty}d\epsilon_{X}\frac{dN_X}
{d\epsilon_X}\sigma_p,
\end{equation}
where $d\epsilon_{X}\cdot dN_X/d\epsilon_{X}$ is the $X$-ray number
density between energies $m_ec^2\epsilon_{X}$ and
$m_ec^2(\epsilon_X+d\epsilon_X)$, $\cos^{-1}\mu_c$ is the collision angle
between a $X$-ray and a $\gamma$-ray, $m_ec^2\epsilon_{th}=2m_ec^2/(1-\mu_c)
\epsilon_{\gamma}$ is the threshold $X$-ray energy for the  pair creation,
and  $\sigma_p$ is the pair creation cross-section, which is given by
\begin{equation}
\sigma_{p}(\epsilon_{\gamma},\epsilon_{X})=\frac{3}{16}
\sigma_{T}(1-v^2)\left[(3-v^4)\ln\frac{1+v}{1-v}-2v(2-v^2)\right],
\label{cross}
\end{equation}
where 
\[
v(\epsilon_{\gamma},\epsilon_{X})=\sqrt{1-\frac{2}{1-\mu_c}\frac{1}
{\epsilon_{\gamma}\epsilon_{X}}},
\]
and  $\sigma_{T}$ is the Thomson cross section.

In this paper, we take the blackbody radiation from the star as the $X$-ray
field. At the distance $r$, the photon number density
 is given by the Planck law,
\begin{equation}
\frac{dN_X}{d\epsilon_X}=\frac{1}{4\pi}\left(\frac{2\pi m_ec^2}{ch}\right)^3
\left(\frac{A_s}{4\pi r^2}\right)
\frac{\epsilon_X^2}{\exp(m_ec^2\epsilon_X/kT_s)-1},
\label{soft}
\end{equation}
where $A_s$ is the emitting area, and $kT_s$ refers to the surface
temperature. For the values of $A_s$ and $T_s$, observed ones may be used.
\subsection{Pair creation model}
\label{boundary}
\begin{figure}
\begin{center}
\includegraphics[width=8cm, height=7cm]{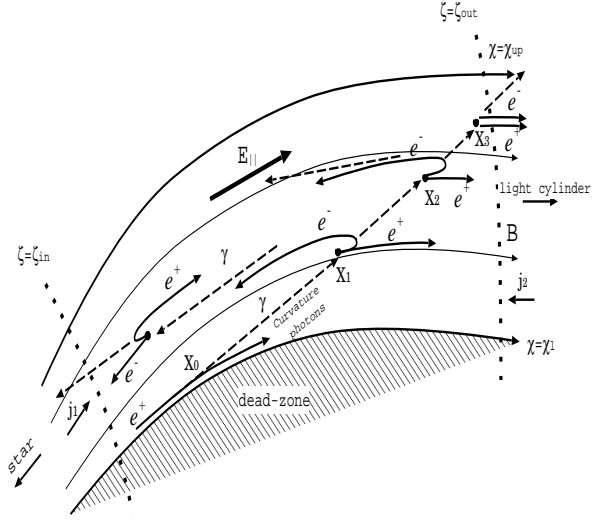}
\caption{The pair creation cascade model in the meridional plane. 
The $\gamma$-rays are radiated by the curvature 
process, and are beamed in the direction of local magnetic field. 
The radiated $\gamma$-rays may convert into the pairs by the 
pair creation process.The $\chi_{l}$, $\chi_{up}$, $\zeta_{in}$, and 
$\zeta_{out}$ present the lower, upper, inner, and outer boundaries of the 
gap, respectively.}
\label{pmodel}
\end{center}
\end{figure}
\label{monte}
We calculate $S(\mathbf{r}$) by  
simulating the pair creation process with Monte Carlo method. 

 Fig.\ref{pmodel} shows the  pair creation cascade model schematically. 
Accelerated particles emit $\gamma$-rays tangentially to the local 
field lines by the curvature process.  
The $\gamma$-rays deviate from the field line on which they are emitted.  
A small fraction of the emitted  $\gamma$-rays may  collide with the $X$-ray 
photons to materialize as electron-positron pairs.
 For example, $G_0$ photons emitted at position $\mathbf{X}_0$ 
in Fig.\ref{pmodel} materialize as  
$\delta N_1$, $\delta N_2\cdots$ ($\ll G_0)$ pairs at positions  
$\mathbf{X}_1, \mathbf{X}_2, \cdots$, respectively. 
The new born electron-positron pairs such as the pairs depicted at the 
position $\mathbf{X}_1$ and $\mathbf{X}_2$, will be separated 
immediately by the field-aligned electric field after their birth. 
The separated pairs also radiate the $\gamma$-rays which become 
new seeds of a sequence of pair creations. The separated pairs 
carry a current, and provide a space charge to screen out the electric field. 

On the other hand, as would be expected near the upper boundary of the gap, 
if the potential drop along the field lines is too small to separate 
a pair, then the pair would go out the  gap together, such as pair 
depicted at position $\mathbf{X}_3$.
 This kind of  pairs does not affect the electrodynamics in the gap, 
in spite of the fact that they are  produced in the gap.
 The pairs produced  outside of the gap do not affect the dynamics either, 
 because they do not directly come into the gap. 
 
To obtain the source term $S(\mathbf{r})$ at each position, we calculate  
the annihilated photon number $\delta N$ at the position $\mathbf{r}$. 
In terms of mean free path of a $\gamma$-ray with energy
 $\epsilon_{\gamma}$,  $l_p(\mathbf{r},\epsilon_{\gamma})
=c/\eta_{p}(\mathbf{r},\epsilon_{\gamma})$, the number of annihilated 
$\gamma$-rays (produced pairs) between the 
distances $s$ and $s+ds$ from the emission point ($s=0$) is
\begin{equation}
\delta N=\frac{G_0}{l_p}\exp\left(-\int_0^s\frac{1}{l_p}ds'\right)ds.
\label{aninum}
\end{equation}  
For the typical pulsar parameters,  the mean free path $l_p$ is found to be 
much longer than the light radius, 
$l_p\gg\varpi_{lc}$, so that we can approximate equation (\ref{aninum}) as 
$\delta N\sim (G_0/l_p)ds$. 
We  assume  that $l_p$ does not depend on the position, taking 
the value at $r=0.5\varpi_{lc}$ with the collision angle $\pi/2$ ($\mu_c=0$). 
This assumption is probably a rough treatment because the $X$-ray number 
density  (\ref{soft}) decreases  with the distance from the centre of the star.
 In the present model, the pair creation position is given by  uniform 
 random number for the given element of the $\gamma$-ray flux. It is notable
 that  the $\gamma$-rays are beamed in the direction 
of local particle's motion 
$\mathbf{v}=\mathbf{v}_{||}+v_{co}\mathbf{e}_{\phi}$, 
so  that an  angle between the direction of the propagation and the 
meridional plane is $\theta_{\gamma}=\cos^{-1}[\sqrt{1-(\Omega\varpi
/\varpi_{lc})^2}]$. Hence, the displacement in the meridional plane is 
reduced to be $s\cos\theta_{\gamma}$, where $s$ is the three-dimensional path 
length of the $\gamma$-rays.

Using the above pair creation model, the source term $S(\mathbf{r})$ is 
obtained by the following procedure (see also Fig.\ref{pmodel}). 1) 
We calculate the number $G_0$ of the emitted photons  per 
unit time per unit volume at the position $\mathbf{X}_0$. 
2) We calculate the number of  photons converted into pairs in a distance 
 $s_0(\sim\varpi_{lc})$  by $\delta G=G_0[1-\exp(-\int_0^{s_0}1/l_pds')]$. 
3) We divide the $\delta G$ photons into $n$ flux elements, and pick up $n$ 
points on the three-dimensional path length 
($s_i<s_0, i=1,\cdots n$) randomly. 
The pair creation point $\mathbf{X}_{i}$ in the meridional plane are 
deduced from $|\mathbf{X}_i-\mathbf{X}_0|=s_i\cos\theta_{\gamma}$. 4) We 
calculate the produced pairs per unit time per unit volume as the source 
term $S(\mathbf{X}_i)$.  If potential drop between a pair creation point
 and  boundaries is small ($e\delta\Phi_{nco}< 10^3m_ec^2$ 
in this paper), we do not count the pair as the source. The emitting points 
of the $\gamma$-rays are scanned along the field line 
(from the inner boundary to the outer boundary) and across the magnetic 
 field lines (from the lower boundary to the upper boundary). 
This direction of  scanning across the field lines has an advantage
 because the $\gamma$-rays propagate always in convex side of the field lines. 
 The particle's number density along the magnetic 
field line follows the equation (\ref{contin}). For the simplest treatment, 
all the curvature photons are assumed to be emitted at the critical energy, 
\begin{equation}
E_c=\frac{3}{4\pi}\frac{hc\Gamma^3}{R_c}=0.1\left(\frac{\Omega}
{\mathrm{100s^{-1}}}\right)\left(\frac{\Gamma}{10^7}
\right)^3\left(\frac{R_{c}}{\varpi_{lc}}\right)^{-1} \ \ \mathrm{ GeV},
\end{equation}
with the rate, 
\begin{equation}
P=\frac{8\pi}{9}\frac{e^2\Gamma}{hR_c}=3.2\times10^6\left(\frac{\Gamma}{10^7}
\right)\left(\frac{R_{c}}{\varpi_{lc}}\right)^{-1}\ \ \mathrm{s^{-1}}.
\label{photons}
\end{equation}

 We introduce the conventional dimensionless variables; 
\begin{equation}
\tilde{\mathbf{r}}=\frac{\omega_p}{c}\mathbf{r},  \ \ \ \
\omega_p=\sqrt{\frac{4\pi e^2}{m_e}\frac{\Omega B_{lc}}{2\pi ce}},
\end{equation}
\begin{equation}
\tilde{\Phi}_{nco}=\frac{e}{m_ec^2}\Phi_{nco}, \ \ \ \
\tilde{\mathbf{E}}=-\tilde{\nabla}\tilde{\Phi}_{nco},
\end{equation}
\begin{equation}
j_{\pm}=\frac{ev_{||\pm}N_{\pm}}{\Omega B/2\pi}, \ \ \ \ \tilde{\mathbf{B}}
=\mathbf{B}/B_{lc},
\end{equation}
 and 
\begin{equation}
\mathbf{\beta}=\frac{\mathbf{v}}{c},
\end{equation}
where $B_{lc}$ is the strength of the magnetic field at the point on which the 
last open field line is tangent to the light cylinder. 
By using dimensionless charge density, 
\begin{equation}
\tilde{\rho}=\frac{\rho}{\Omega B_{lc}/2\pi c}=\frac{\tilde{B}}{\beta_{||}}
(j_+-j_-), \ \ \ \ \tilde{\rho}_{GJ}=-\tilde{B}_z,
\label{lcharge}
\end{equation}
the Poisson equation (\ref{poisson}) and the continuity equation (\ref{contin})
 are rewritten as 
\begin{equation}
\tilde{\triangle}\tilde{\Phi}_{nco}=-[\frac{\tilde{B}}{\beta_{||}}(j_+-j_-)
+\tilde{B}_z],
\end{equation}
and
\begin{equation}
\tilde{\mathbf{B}}\cdot\tilde{\nabla}j_{\pm}=\pm\tilde{S}(\tilde{\mathbf{r}}),
\label{nconeq}
\end{equation}
respectively.
\subsection{Boundary conditions and method of calculation}
\label{method}
We assume that the lower and the upper boundaries are laid on 
the magnetic surfaces labeled by  $\chi_{l}$ and $\chi_{up}$, respectively.
The inner and the outer boundaries are defined by the surface on which 
$\tilde{E}_{||}$  vanishes,
\begin{equation}
\tilde{E_{||}}(\zeta_{in})=\tilde{E_{||}}(\zeta_{out})=0,
\label{cond1}
\end{equation}
where $\zeta_{in}$ and $\zeta_{out}$, which are in general function of $\chi$, 
 denote the location of the inner and the outer boundaries, 
respectively (Fig.\ref{pmodel}).

The conventional polar cap and outer gap models have their current 
densities flowing in opposite directions to  each other. For this difference, 
it is impossible that these two acceleration regions
are located on the  same magnetic field lines. Therefore, we anticipate
 that the inner, the upper and the lower boundaries are directly linked with 
the star without any potential drop. We then impose
\begin{equation}
\tilde{\Phi}_{nco}(\zeta_{in})=\tilde{\Phi}_{nco}(\chi_{up})=
\tilde{\Phi}_{nco}(\chi_{l})=0.
\label{cond2}
\end{equation}
The conditions (\ref{cond1}) and (\ref{cond2}) are not satisfied on
arbitrary given boundaries, because both  the Dirichlet-type and the
Neumann-type conditions are imposed on the inner boundary $\zeta_{in}(\chi)$.
 We, therefore,  solve the location of the inner boundary.
 By moving the inner boundary
step by step iteratively, we seek for the boundary that satisfies the
required conditions.

We have to consider the  stability  of the gap. 
The stability condition for the inner boundary is    
$\partial\tilde{E}_{||}/\partial \tilde{s}_{||}\ge0$ 
 such that external  electrons may not freely come into the gap 
through the inner boundary. This is similar to the force-free surface in the 
quiet aligned model by Michel (1979). In the same way, for the outer boundary 
it is  $\partial\tilde{E}_{||}/\partial \tilde{s}_{||}\le0$. Furthermore, 
we postulate that the gap is unstable if the field-aligned electric field 
changes its direction. In other word, the sign of the 
field-aligned electric field is positive definite, as it guarantee 
that the positrons and the electrons propagate outwardly and inwardly,
 respectively. For the non-vacuum case, if super-GJ ($|\rho|>|\rho_{GJ}|$) 
region is dominant in the gap, then the electric field may change its 
direction.

The continuity equation (\ref{nconeq}) satisfies conservation law of 
the longitudinal current density, that is, 
\begin{equation}
j_{tot}\equiv j_++j_-=\textrm{const along field line},
\label{cons}
\end{equation}
where $j_{tot}$ is the total current density in units of the GJ
 value. The total current is made up  of the  three components, 
\begin{enumerate}
\renewcommand{\theenumi}{(\arabic{enumi})}
\item $j_g$, the current carried by the electrons and the positrons  produced 
in the gap, 
\item $j_1$, the current carried by the positrons (e.g. originating in
  sparking on the stellar surface) coming into the gap through 
the inner boundary,
\item $j_2$,  the current carried by the electrons (e.g. originating in 
 pulsar wind region) coming into the gap through the outer boundary.
\end{enumerate}
We cannot deduce the value of $j_{tot}$ from any  gap model, because the 
current circulates in the magnetosphere globally. 
The outer gap, the polar cap and the pulsar wind may  interact each other 
through the current, so that the current should be determined by some global 
conditions. In our local model, therefore, we regard  
the values of $(j_g,j_1,j_2)$ as a set of free parameters. The boundary
 conditions on the current at the inner 
and the outer boundaries are given by
\begin{equation}
j_{+}(\zeta_{in})=j_1(\chi),
\end{equation}
and
\begin{equation}
j_-(\zeta_{out})=j_2(\chi),
\end{equation}
respectively. In terms of  ($j_g,j_1,j_2$), 
the conservation low (\ref{cons}) yields 
\begin{equation}
j_1+j_2+j_g=j_{tot}.
\end{equation}

As has been mentioned, any local model should have some degrees of freedom 
because the model includes global quantities such as the electric current. The 
free parameters cannot be eliminated unless one solves the interaction with 
other parts of the magnetosphere (e.g. with pulsar wind). Furthermore, 
we have the free boundary problem with both  the Dirichlet-type and the
Neumann-type conditions. Hence, 
choice of a set of free model parameters is somewhat arbitrary. For 
convenience for the iteration, we choose the model parameters as follows. The 
external current sources ($j_1,j_2$) are obviously suitable to be free 
(e.g., in the traditional gap models, one assumes $j_1=j_2=0$). The lower 
boundary would be chosen as a model parameter, but it is traditionally set 
to be the last open field line, that is, $\chi_{l}=1$. We give also the 
locations of the upper boundary $\chi_{up}$ and the outer boundary 
$\zeta_{out}(\chi)$, thereby we obtain a specific outer gap, namely we solve 
$j_g(\chi)$ and the location of the inner boundary $\zeta_{in}(\chi)$. Giving 
different  $\zeta_{out}(\chi)$ and $\chi_{up}$, we obtain 
 a series of models for which  $j_g(\chi)$ and $\zeta_{in}(\chi)$ are obtained.

One specific solution is determined by iteratively as follows. 
1) By assuming  the field-aligned electric 
field $E_{||}$  and the position 
of inner boundary  $\zeta_{in}(\chi)$, we start our calculation. 2) We calculate 
the particle's Lorentz factor (\ref{lorent}) from $E_{||}$. 
3) We solve the curvature radiation and the pair creation processes by the 
method described in \S\S\ref{monte}, and then obtain the particle's number 
density in the gap. 4) Having the space charge density, we solve 
the Poisson equation to obtain  new field-aligned electric field  
and the inner boundary that 
satisfies the required conditions.  5) We repeat 2)-4) until all the physical 
quantities are self-consistent. 
 
The values of all the physical quantities go quickly toward 
convergence through the iteration. For example, if the field-aligned 
electric field is over estimated, then the resultant particle's number 
(or charge) density calculated with the radiation and the pair creation 
processes becomes larger than the solution. With this over estimated number
 density, however, the Poisson equation gives a smaller electric field in 
the next step.  
\section{Results}
\label{result}
\subsection{Model parameters}
\label{parameter}
As described in the last section, we take  ($j_1,j_2$), 
($\chi_{l},\chi_{up}$) and $\zeta_{out}$ as the model parameters.  
We fix $\chi_{l}=1$ (i.e. the lower boundary is the last open field line), 
and $\zeta_{out}=1.6$ such that the trans-field path length  
of $\gamma$-rays in the gap may becomes significant. To examine the  effect 
of variation in the external current, we consider the following four cases;
\begin{enumerate}
\renewcommand{\theenumi}{Case~\arabic{enumi}:}
\item $(j_1,j_2)=(0.0,0.0)$ for $\chi_{up}\geq \chi> \chi'$, 
  but $(j_1,j_2)=(10^{-5},10^{-5})$ for $\chi'\geq \chi\geq \chi_{l}$, where
 $\chi'=\chi_{l}+0.02(\chi_{up}-\chi_{l})$,
\item $(j_1,j_2)=(0.1,0.1)$ for $\chi_{up}\ge \chi\ge \chi_{l}$, 
\item $(j_1,j_2)=(0.2,0.0)$ for $\chi_{up}\ge \chi\ge \chi_{l}$,
\item $(j_1,j_2)=(0.0,0.2)$ for $\chi_{up}\ge \chi\ge \chi_{l}$.
\end{enumerate}
In Case~2, the positronic and the electronic currents of 10\% of 
the typical GJ value are injected into the gap from the inner and the outer 
boundaries, respectively. In Case~3 (or Case~4), on the other hand, 
the only positronic (or electronic) current comes into the gap through 
either side of the gap. In Case~1, we consider the case of no external current,
 that is, the whole  current is carried  by the  pairs produced in the
 gap. Because we need seed $\gamma$-rays 
to start the pair creation cascade in Case~1, we assume  a very little 
currents ($j_1=j_2=10^{-5}$) in a thin layer (2\% of the trans-field thickness)
 just above the last open field line.

In general, $j_g$ increases with $\chi_{up}$ (i.e. with the trans-field 
thickness).  But, if  $j_g$ becomes larger
 than a critical value,  we cannot get the stable 
solutions because the field-aligned electric field changes 
its sign in the gap.  In this paper,  we regard the maximized $\chi_{up}$ as a 
natural boundary, and thereby we obtain $\chi_{up}\sim1.22$ (Case~1),
 $\sim1.2$ (Case~2), $\sim 1.13$ (Case~3) and $\sim1.21$ (Case~4). 
For the pulsar parameters, 
we have used the values of the  Vela pulsar; $\Omega=70.6\mathrm{s^{-1}}$, 
$B_s=6.68\cdot10^{12}$G, $kT_s=150$eV, and $A_s/A_{*}=0.066$ 
(\"{O}gelman, Finley \& Zimmerman 1993), where $B_s$ is the polar field 
strength, and $A_*$ is the area of the stellar surface. 

\subsection{Vacuum gap}
Before we investigate the electrodynamics of the non-vacuum gap, 
let us see some properties of the vacuum solution.

If the gap width $W_{||}$ along the magnetic field lines is much shorter than 
the trans-field thickness $D_{\perp}$, then the one-dimensional 
approximation works out well. From the one-dimensional Poisson equation, 
$d\tilde{E_{||}}/d\tilde{s}_{||}=\tilde{B}_z$, the inner and the outer 
boundaries satisfy
\begin{equation}
\int_{\tilde{s}_{||in}}^{\tilde{s}_{||out}}\tilde{B}_zd\tilde{s}_{||}=0,
\label{varho}
\end{equation}
where $\tilde{B}_z$ is the dimensionless GJ charge density, 
 $\tilde{s}_{||in}$ and  $\tilde{s}_{||out}$ are  the locations of the 
inner and the outer boundaries, respectively,  in terms of the arc length 
along the field line from the stellar surface. In this case, 
the field-aligned electric field $E_{||}$  varies quadratically along the 
field lines such as one displayed in Fig.\ref{wedgap}.

\begin{figure}
\begin{center}
\includegraphics[width=14cm, height=6cm]{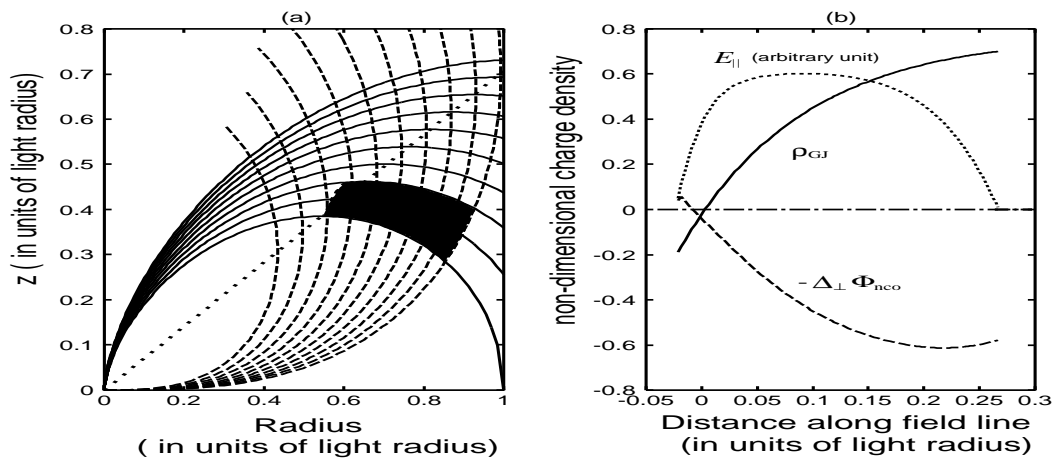}
\caption{A solution of the vacuum gap. (a) The gap in the magnetosphere 
(filled region). The solid and dashed lines show the dipole field lines and 
the curve lines that perpendicular to the field lines, respectively. The 
dotted line is the null surface. (b) The solid, dashed, and dotted lines show 
the distributions of the GJ density, the two-dimensional term, and 
acceleration field, respectively, along a magnetic field line.}
\label{vagap}
\end{center}
\end{figure}

CHR pointed out that the trans-field thickness is 
suppressed by the pair creation, and postulated that the gap width 
is much longer the trans-field thickness, $W_{||}\gg D_{\perp}$. 
By solving the Poisson equation with this slab-like geometry, 
they found that the inner boundary is located close to the  null surface. 
While CHR solve the Poisson equation with a recti-linear
geometry, the present model gives  a vacuum solution with the dipole
 magnetic field in Fig.\ref{vagap}(a).  The result retains a property of CHR; 
that is, \textit{the inner boundary is located near              
the null surface}. This property does not depend on the location of the 
outer boundary.

In the elongated gap case, instead of the equation (\ref{varho}), 
 the inner and the outer boundaries satisfy 
\begin{equation}
\int_{\tilde{s}_{in}}^{\tilde{s}_{out}}(\tilde{\triangle}_{\perp}
\tilde{\Phi}_{nco}+\tilde{B}_z)d\tilde{s}_{||}=0,
\end{equation}
where $\tilde{\triangle}_{\perp}$ is the trans-field part of the Laplacian. 
Fig.\ref{vagap}(b) shows the distribution of $\tilde{B}_z$,
  the transverse term $\tilde{\triangle}_{\perp}\tilde{\Phi}_{nco}$, and 
 $E_{||}$ on a magnetic field. The transverse term represents the effect of 
surface charge on the upper and the lower boundaries. From Fig.\ref{vagap}(b),
 we find that the GJ charge density and the transverse term almost balance out
 in most part of the gap, $\tilde{B}_{z}\sim-\tilde{\triangle}_{\perp}
\tilde{\Phi}_{nco}$.  Therefore, the distribution of $E_{||}$ has a plateau in
 the  middle of the gap as the dotted line in Fig.\ref{vagap}(b) shows. 
\subsection{Non-vacuum gap}
\label{nonvacuum}
In this subsection, we discuss the electrodynamics of the non-vacuum gap. 
The lowest order of the Fourier components of the two-dimensional term is 
represented by $\tilde{\triangle}_{\perp}\tilde{\Phi}_{nco}\sim-
\tilde{\Phi}_{nco}/\tilde{D}_{\perp}$, and  therefore the two-dimensional term
 plays the role of a  $positive$ charge density in the gap. 
\subsubsection{Electric field structure}
\label{efs}
\begin{figure}
\begin{center}
\includegraphics[width=10cm, height=8cm]{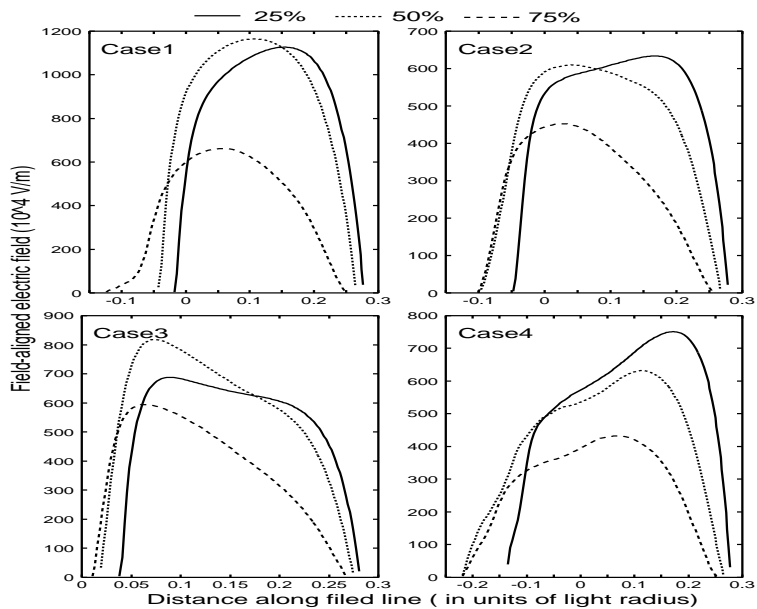}
\caption{The electric field structure for Case~1, Case~2, Case~3, and Case~4.
 The solid, dotted, and dashed lines show the distributions of
the field-aligned electric field on the three different magnetic field lines, 
which locate, respectively, 25, 50, and 75\% of the gap trans-field thickness
 measured from the last open field line. The abscissa represent the arc length
 with the origin at the null surface.}
\label{electric}
\end{center}
\end{figure}

Fig.\ref{electric} shows the distributions of the field-aligned electric field 
$E_{||}$ along the magnetic field lines. The solid, dotted and dashed lines
 represent, respectively, the electric  field on the magnetic field lines 
threading the gap at  25, 50, and 75 \% of the  trans-field thickness 
from the lower boundary.  The abscissa refers to the arc length from the null
 surface. The positive and the negative values indicate outside and inside, 
respectively, with respect to the  null surface.

The variations of $E_{||}$ along the magnetic field line 
 for all cases are similar to that  for the vacuum case displayed in 
Fig.\ref{vagap}. In these cases, because the trans-field thickness is smaller  
than the gap width, the transverse term $\tilde{\triangle}_{\perp}\tilde{\Phi}_{nco}$ is  important. The value of $E_{||}$ is determined in such a way that 
$\tilde{\triangle}_{\perp}\tilde{\Phi}_{nco}$, 
the local charge density $\tilde{\rho}$ and the GJ charge density 
$-\tilde{B}_z$ almost balance out in most part of the gap, 
$-\tilde{\triangle}_{\perp}\tilde{\Phi}_{nco}-\tilde{\rho}\sim\tilde{B}_z$
 (see Fig.\ref{charge}).  
As a result, the distribution of the electric field shows a plateau in 
contrast with the quadratic shape (such as Fig.\ref{wedgap}) in the 
one-dimensional model, in which the transverse term is ignored.
  
The effect of the external  currents ($j_1,j_2$) 
produces the difference in the distribution of  $E_{||}$. For example, 
 $E_{||}$ on the plateau in Case~3 gradually decreases with the distance, 
while that in Case~4 gradually increases. In terms of  $j_1$ and $j_2$, the local charge
 density (\ref{lcharge}) is  rewritten  as
 $\tilde{\rho}=\tilde{B}(j_1-j_2-j_g+2\delta j)/\beta_{||}$,
 where $\delta j$ represents the current density carried by the positrons 
produced between the inner boundary and the position considered (e.g. 
$\delta j=0$ at the inner boundary, and $\delta j=j_g$ at the outer boundary).
 For Case~3 ($j_1\neq0, j_2=0$), the charge density 
at the outer boundary is $\tilde{B}(j_1+j_g)/\beta_{||}$, indicating that 
the external positronic current $j_1$  assists the current $j_g$ 
in screening of $E_{||}$ at the outer boundary. Therefore, the strength 
of $E_{||}$  near the outer boundary is smaller than that near the inner
 boundary. In the same way, the external electric current $j_2$ for 
Case~4 assists the  screening of $E_{||}$ at the inner boundary, where the 
charge density is $\tilde{B}(-j_2-j_g)/\beta_{||}$. Hence the strength of 
$E_{||}$ near the inner boundary is larger than  that near the outer boundary.
 To be  expected, if both the positronic and the electronic currents are 
injected the respective boundaries such as Case~2, these two contributions 
in the Poisson equation cancel each other, so that the plateau becomes 
almost flat.  

Another effect of the external currents $(j_1,j_2)$ appears in the 
position of the inner boundary.  
The inner boundary for  Case~3 is located outside of the null surface, and 
as a result the gap does not include the null surface,  
although the boundaries for the other cases are located  
inside of the null surface. This difference is discussed in \S\S\ref{location}.
\subsubsection{Current density}
\begin{figure}
\begin{center}
\includegraphics[width=10cm, height=8cm]{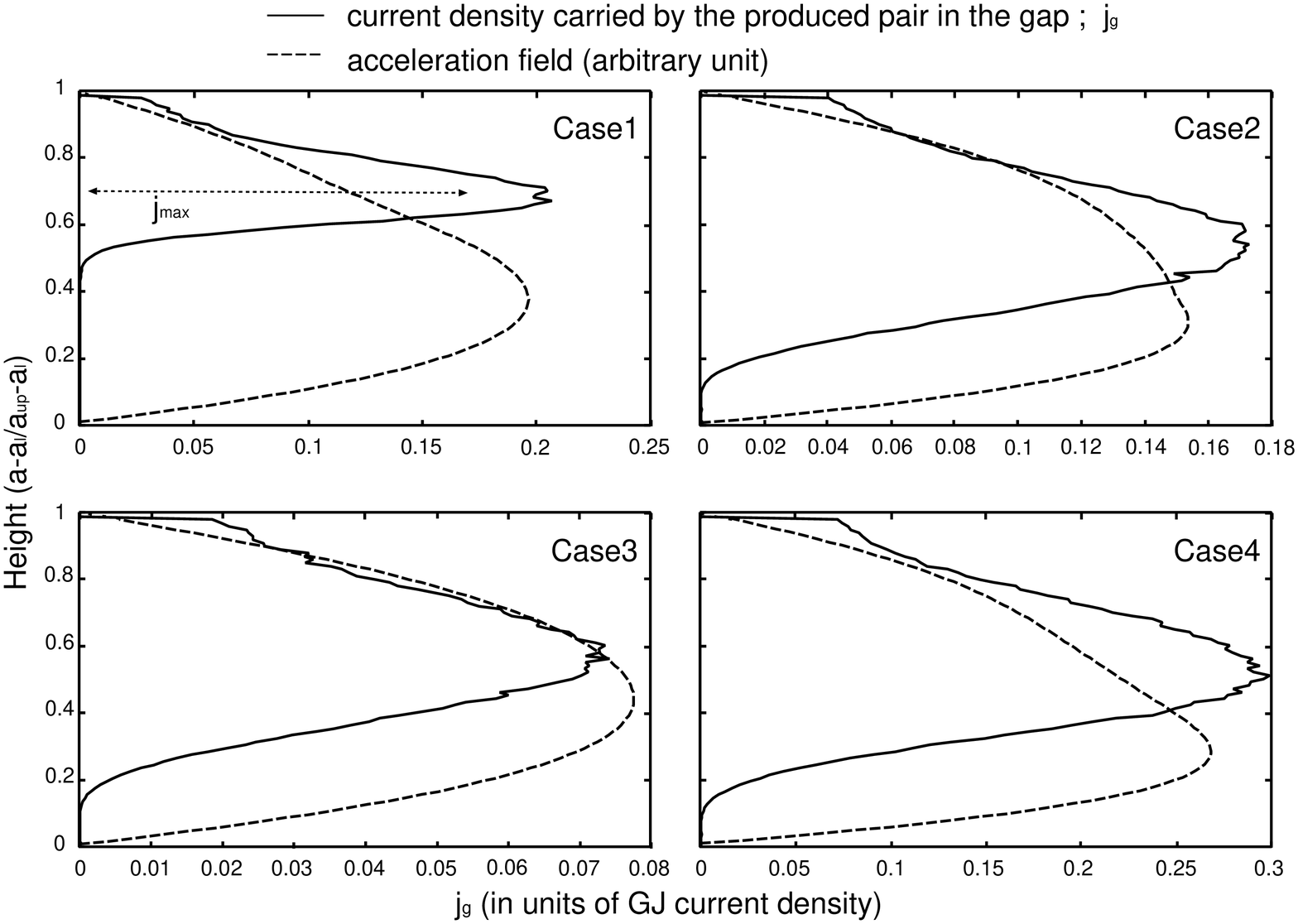}
\caption{The trans-field structures of the current density and the 
field-aligned electric field. The solid line shows the trans-field distribution
 of the current $j_g$ carried by the pairs produced in the gap. The dashed 
line shows the maximum value (arbitrary units) of $E_{||}$ on each field lines.
}
\label{current}
\end{center}
\end{figure}

The trans-field structures found in the current density and the field-aligned 
electric field $E_{||}$ are summarized in Fig.\ref{current}. 
The solid lines show the distribution of the current $j_g$  across the field 
lines. The  maximum value (arbitrary units) of 
 $E_{||}$  on each field line is also plotted in Fig.\ref{current}.
  
We find that maximum value of $j_g$ for each case is 
$j_{max}\sim0.2$ (Case~1), $\sim0.17$ (Case~2), 
$\sim0.07$ (Case~3) and $\sim0.3$ (Case~4).
The total current density on each magnetic field line is represented by 
$j_{tot}=j_g+j_1+j_2$. In Case~1, for example, 
 because no particles come into the gap through both the boundaries,
 (i.e., $j_1=j_2=0$), the total current is carried only by the pairs produced 
in the gap, that is, $j_{tot}=j_g$. The total current densities at the maximum
 are $j_{tot}=j_{max}\sim0.2$ (Case~1), $j_{max}+j_1+j_2\sim 0.37$ (Case~2), 
$j_{max}+j_1\sim0.17$ (Case~3), and $j_{max}+j_2\sim0.4$ (Case~4).

\begin{figure}
\begin{center}
\includegraphics[width=7cm, height=7cm]{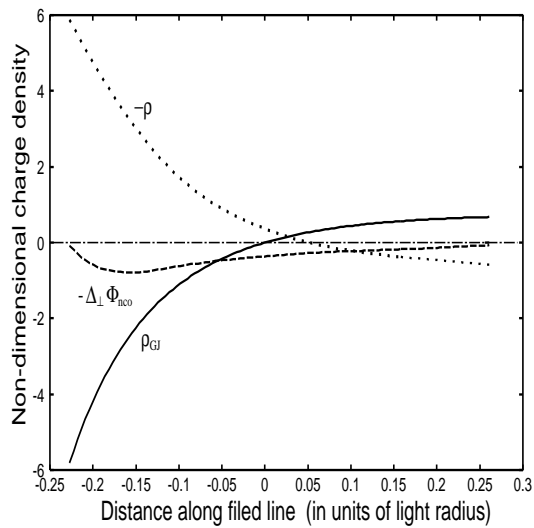}
\caption{The distributions of the GJ charge density (solid line), 
the transverse term (dashed line) and the space charge density (dotted line) 
 on the magnetic field line, on  which $j_g\sim j_{max}$ for Case~4.}
\label{charge}
\end{center}
\end{figure}

As mentioned in \S\S\ref{parameter}, the solution disappears 
if $j_g$ exceeds the critical value because the super-GJ region becomes 
dominant in the gap so that the field-aligned electric field changes 
its direction.  However, it is noteworthy that 
the values $j_g$ for Case~1, Case~2 and especially Case~4 exceed the 
critical value ($\sim0.1$) of the previous one-dimensional model. 
As an effect of the trans-field structure, the transverse term 
$\tilde{\triangle}_{\perp}\tilde{\Phi}_{nco}$ behaves as 
the positive charge.  Therefore, if the super-GJ points appear in the 
negatively charged region, the effect of it in the Poisson equation can be
 reduced by the transverse term.  In  Case~4, for example,   
the super-GJ points tends to  appear in the negatively charged region around 
the inner boundary due to the external electric current $j_2$ 
(see Fig.\ref{charge}). Then, we find that the shape of 
its inner boundary (see next subsection) and the gap structure are formed 
so that the transverse term cancels the super-GJ points as 
Fig.\ref{charge} shows. Because we increase $j_g$ until 
the transverse term marginally sustain the super-GJ region, we obtain 
the critical value ($\sim0.3$) of Case~4 increasing significantly 
from the one-dimensional one ($\sim0.1$). The situation such as Case~4 
 is very important for obtaining the solution of the increased current. 
On the contrary, if the super-GJ space charge density appears near the outer 
boundary, it is  positive, so that the transverse term does not reduce 
the super-GJ charge density but the opposite. In such a case (e.g. Case~3 with 
positive external current), the critical current density is rather small.

In the outer gap model by CHR, the pairs are supposed to be created  
much in the convex side (i.e. in the upper half), and a large current 
flows there, while they assumed  the vacuum region in the lower half. 
This type of gap is realized in Case~1 by assuming no external currents.
 From the panel for Case~1 in Fig.\ref{current}, 
we find that the vacuum-like region occupies a lower half of the gap,
 and the current flowing (non-vacuum) region is restricted 
in the upper half region. As far as the trans-field structure is concerned,
 our electrodynamical model reproduces the structure similar to that assumed 
in CHR. However, the location of the inner boundary manifest itself 
in different manner, as shall be shown in the next subsection. 

 Fig.\ref{current} shows that the current $j_g$ increases with increasing 
height, and then decreases. The increase is simply because 
the $\gamma$-rays propagate the convex side of the field lines and make pairs. 
The decrease is caused by the following 
effects. Firstly, the upper part is illuminated mainly by the $\gamma$-rays 
which were radiated in the middle region, while almost of
 the $\gamma$-rays emitted at the lower half part  escape through 
the inner or the outer boundaries before reaching the upper region.  
Secondly, the field-aligned 
electric fields, the Lorentz factor ($\propto E_{||}^{1/4}$) 
 and in turn the energy of the emitted  $\gamma$-rays 
($\propto E_{||}^{3/4}$) 
 decrease with increasing of the  height in the upper half region.
 Furthermore, the threshold energy of the $X$-rays colliding with the 
low-energy  $\gamma$-rays moves to the Wien region of the Planck law, 
and thereby, the mean free path of 
the $\gamma$-rays sharply increase with decreasing of its energy.  
By combinations of these effects, the current density
 $j_g$  decreased with increasing of the height in the upper half region. 

In contrast, the exponential increase of the currents near the upper boundary 
are predicted in CHR. 
This case may be obtained, if one takes non-thermal $X$-ray fields with 
a modest power-low index such as the Crab pulsar, or if one considers 
a case in which  the strength of $E_{||}$ is ineffective to the Lorentz 
factor and the energy of 
the $\gamma$-rays (e.g. the unsaturated motion of particles). 
These possibilities are investigated in the subsequent papers.
\begin{figure}
\begin{center}
\includegraphics[width=10cm, height=8cm]{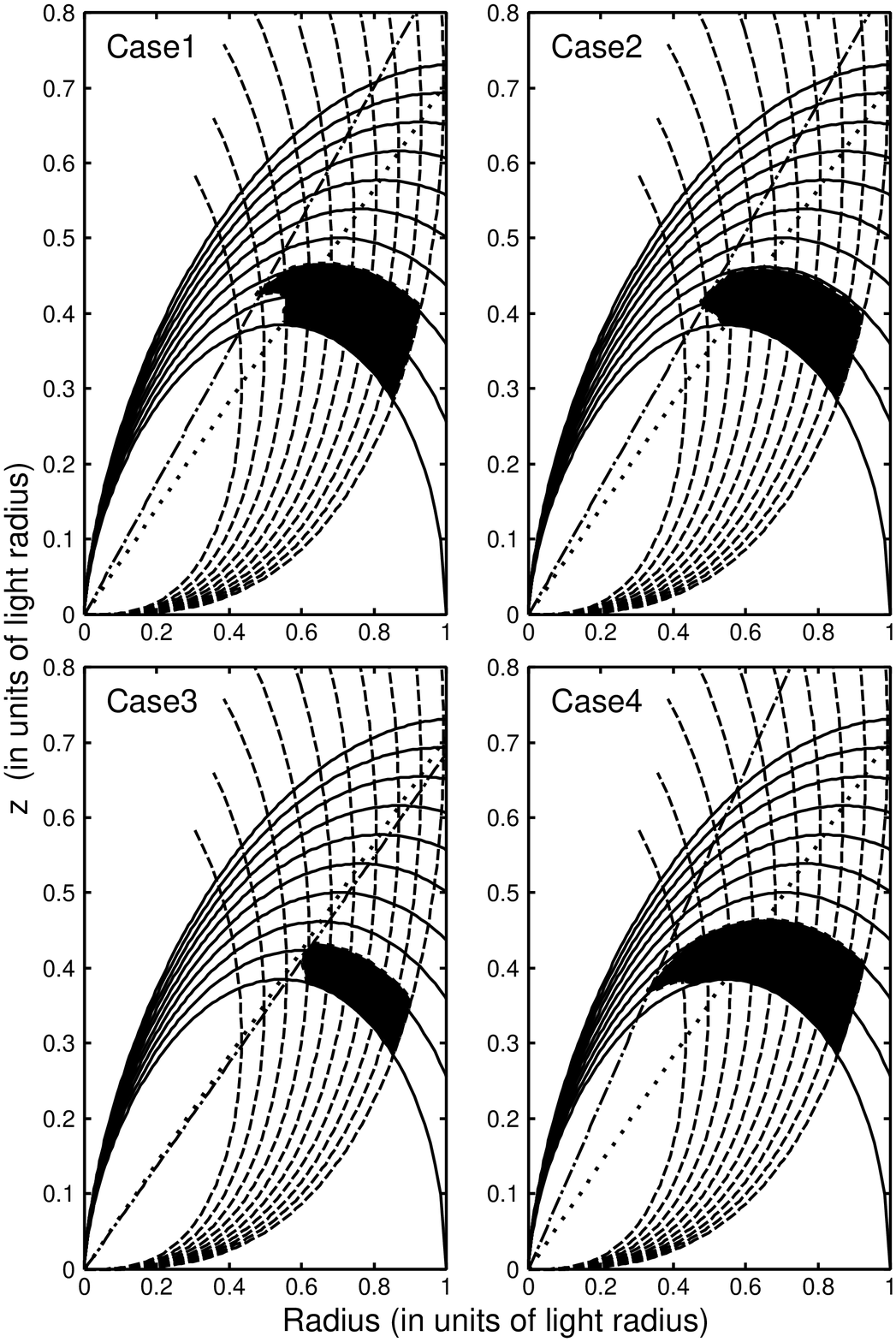}
\caption{The geometry of the outer gap (filled region in each panel). 
The solid 
and dashed, and dotted lines are the same as Fig.\ref{vagap}. On the 
dashed-dotted line in each panel, the condition, 
$j_{max}+j_2-j_1=\tilde{B}_z/\tilde{B}$, is satisfied 
(see text).}
\label{nvagap}
\end{center}
\end{figure}

\subsubsection{Location of the inner boundary} 
\label{location}
The filled regions in Fig.\ref{nvagap} show the shape of the gap. 
The $beak$-like 
shape of the 
inner boundary is remarkable when one compares it with that of 
the vacuum case. The magnetic field line on which the 
current density $j_g$ is maximized at $j_{max}$ runs through the cusp of 
the beak.  

An important feature of the external currents ($j_1,j_2$) 
appears in the position of  the inner boundary of the lower end part, 
where few  pairs are produced. 
If $j_1$ and $j_2$ are equal such as Case~1 and Case~2,  
the inner boundary is located close to the  null surface. 
If $j_1$ and $j_2$ are not equal such as Case~3 and Case~4, on the other hand, 
the inner boundary shifts outside or inside. 
This is because the effective charge neutral shifts relative to the 
conventional null surface by the space charge caused by the external 
current.  This effect is originally pointed out by  the  one-dimensional 
model of Hirotani \& Shibata (2001).

As mentioned above, 
the inner  boundary at the lower part region of Case~1  comes to close 
the null surface. On the other hand,
 we find that the inner boundary at the cusp, where  $j_g=j_{max}$,  
leaves from the null surface  and moves toward the stellar surface. 
Let us investigate this displacement. From the 
stability condition  discussed  in \S\S\ref{method}, we have
\begin{equation}
\frac{\partial \tilde{E}_{||}}{\partial \tilde{s}_{||}}=
\tilde{\triangle}_{\perp}\tilde{\Phi}_{nco}
+[\tilde{B}(j_+-j_-)/\beta_{||}+\tilde{B}_z]\ge0.
\label{bouncond}
\end{equation}
With the help of  $\tilde{B}_z/\tilde{B}\leq1$ and the fact that   
both $\tilde{\triangle}_{\perp}\tilde{\Phi}_{nco}\sim0$ at the  and $\beta_{||}\sim1$ hold on the cusp, we have
\begin{equation}
(j_{max}+j_2-j_1)\le \tilde{B}_z(\zeta_{in})/\tilde{B}(\zeta_{in})\le 1,
\label{incond}
\end{equation}
where we have used the fact that $j_+-j_-$ becomes $j_1-j_2-j_{max}$ at 
the cusp. This condition indicates that the inner boundary 
is located between the stellar surface where $\tilde{B}_z/\tilde{B}=1$ and 
the  position on which the local charge density caused by the longitudinal 
currents  equals the GJ charge density.

In Case~1 and Case~2, because $j_1$ equals $j_2$, 
the condition (\ref{bouncond}) becomes 
\begin{equation}
j_{max}\leq\tilde{B}_z(\zeta_{in})/\tilde{B}(\zeta_{in})\leq1.
\label{bouncond1}
\end{equation}
The dashed-dotted lines for Case~1 and Case~2 
in Fig.\ref{nvagap} show the positions on which 
the condition $j_{max}=\tilde{B}_z/\tilde{B}$ ($j_{max}=0.2$ 
in Case~1 and $j_{max}=0.17$ in Case~2) is satisfied.
 We find that the cusp is located significantly closer to the position where 
the space charge density, $-j_{max}\tilde{B}$, equals the GJ charge 
density, $-\tilde{B}_z$. Noting that $\tilde{B_z}/\tilde{B}$ increases with 
 decreasing of  $r$, we expect that the cusp  shifts toward stellar 
surface as $j_{max}$ increases. If the current carried by the particles 
produced in the gap were the GJ value (i.e. if $j_{g}=1$), 
the cusp would be located on the stellar surface, and therefore the outer
 gap would start from the stellar surface. However, because we have no 
solution if $j_g$ exceeds a critical value ($\sim 0.2$ for Case~1), 
we do not obtain the solution in which the inner boundary touches the 
stellar surface. 
 
For Case~3 and Case~4, the condition (\ref{incond}) is rewritten 
as  $(j_{max}-j_1)\le\tilde{B}_z(\zeta_{in})/\tilde{B}(\zeta_{in})\le1$ 
and $(j_{max}+j_2)\le\tilde{B}_z(\zeta_{in})/\tilde{B}(\zeta_{in})\le1$, respectively. 
The dashed-dotted lines in the lower panels of Fig.\ref{nvagap} show  
the positions which satisfy the condition 
$(j_{max}-j_1)\tilde{B}=\tilde{B}_z$ for Case~3 and  
$(j_{max}+j_2)\tilde{B}=\tilde{B}_z$ for Case~4. Such as Case~1 and Case~2, 
the cusps in these cases also take the position on which 
the space charge density equals the GJ charge density. Thus, 
we find that the outer gap extends from the point on which the charge 
density caused by the longitudinal current made up of ($j_{max},j_1,j_2$) 
equals the GJ charge density. 

\section{discussion}
\label{disc}
We have studied the electrodynamics of the outer gap including the trans-field
 structure in the meridional  surface.  
We have found that the inner boundary shifts toward 
the stellar surface from the null surface as $j_g$ or $j_2$ increases, 
and forms the beak shape, which was found in the vacuum gap geometry. 
We have shown the specific solution, of which the current density $j_g$ 
carried by the particles produced in the gap becomes significantly 
lager than the upper limit given by the one-dimensional model ($j_g\sim0.1$).  
\subsection{Comparison with previous works}
After  CHR, the traditional shape of the outer gap have
been based on the vacuum model, in which the gap extends between
the null surface and the light cylinder. Although the traditional
outer gap models have been successful in explaining overall features 
of the observed light curves (e.g. double-peaks in one period, and 
the presence of the bridge emission), the model cannot account for 
the presence of the  outer-wing
 emission and of the off-pulse emission from the Crab pulsar. Recently, 
the  \textit{caustic model} by  Dyks \&  Rudak (2003), and 
 Dyks, Harding \& Rudak (2004) reproduces the
observed light curves  better than the traditional outer gap model.
The caustic  model  demanded that the emission region  extends
from the  polar caps to the light cylinder.  But, the reason that
the gap extends  on both sides of the null surface is not evident.
Our two-dimensional electrodynamical model indicates
 that if  a marginal stable condition, 
$(j_{g}+j_2-j_1)\tilde{B}(\zeta_{in})=\tilde{B}_z(\zeta_{in})$ holds, then 
the inner boundary shifts toward the stellar surface as $j_{g}$ and
$j_2$ increase.

We have actually demonstrated that the gap extends on both sides of 
null surface in Case~4 (Fig.\ref{nvagap}), in which the electrons come into 
the gap through the outer boundary. In the global magnetospheric models
(e.g. Mestel et al. 1985, Shibata 1995),  the electrons circulate
the magnetosphere globally and  turn back into the outer gap. 
 Although the observed light curves may be reproduced  better by
assuming the injection of the  electrons at the outer boundary
($j_1=0, j_2\neq0$), it is not certain whether the injection from the outer 
boundary is favored for reproducing the $\gamma$-ray spectrum. This will be 
studied in a subsequent paper. It is notable that, in the 
one-dimensional model, the observed spectra have been reproduced 
by assuming the injected positrons from the inner boundary 
(Hirotani et al. 2003; Takata et al. 2004).

Although the present model is very simple, we must argue that two or three 
dimensional model will open a way to diagnose the current system in the outer
 gap when the observed phase resolved $\gamma$-ray spectra and light curves 
are compared with the model predictions. The external currents  ($j_1,j_2$) strongly 
control the $\gamma$-ray spectrum by changing the electric field strength. 
On the other hand, the current density $j_g$ and its distribution across the 
field lines reflect the pulse shapes through the location of the gap boundary. 
Thus, combining the spectra and the pulse shapes, 
the electrodynamical model enables us to know the 
current system in the outer gap. 
 
\subsection{Outside of the gap}
In this subsection, we discuss the dynamics in  outside of the gap. 
We shall show that coexistence of low energy particles and the high energy
 particles streaming from the gap is required to hold the ideal-MHD condition,
$E_{||}=0$, in the outside of the gap. 
We shall lead this conclusion with \textit{the reductio ad absurdum}.
 
Taking Case~1 ($j_1=j_2=0$), let us consider the region 
between the stellar surface and the inner boundary. We assume 
at the moment that 
\textit{only the electrons escaping from the gap} exist in that region 
and migrate toward the stellar surface. The Poisson equation and the 
energy conservation for the electrons in the one-dimensional form are 
\begin{equation}
\frac{d\tilde{E}_{||}}{d\tilde{s}_{||}}
=-\frac{\Gamma-\Gamma_0}{\tilde{D}^2_{\perp}}
-\frac{j_g\tilde{B}}{\beta_{||}}+\tilde{B}_z,
\label{outside}
\end{equation}
and 
\begin{equation}
\frac{d\Gamma}{d\tilde{s}_{||}}=-\tilde{E}_{||},
\end{equation}
respectively, where  $\tilde{s}_{||}$ is the arc length along the 
field line from the stellar surface, $\Gamma_0$ is the Lorentz factor of the 
particles on the inner boundary, 
and $\Gamma=1/\sqrt{1-\beta_{co}^2-\beta_{||}^2}$.  The first term in 
the right hand side of the equation (\ref{outside}) represents the 
two-dimensional effect.

Near the inner boundary but outside of the gap, 
because $\beta_{||}\sim 1$ and because $\Gamma\sim\Gamma_{0}$, the equation 
(\ref{outside}) can be rewritten 
 as $d\tilde{E}_{||}/d\tilde{s}_{||}=-j_g\tilde{B}+B_z$. Since 
$j_g\tilde{B}\le\tilde{B}_z$ is satisfied, we find 
$d\tilde{E}_{||}/d\tilde{s}_{||}\ge0$, where the equality is satisfied 
on the inner boundary. By noting $E_{||}=0$ on the boundary by the definition, 
the field-aligned electric field has the negative sign in the outside of the gap,
 so that the Lorentz factor of the escaping electrons decreases. 
If the decrease of $\Gamma$ from $\Gamma_0$ becomes a considerable value, 
the transverse term dominates the right hand side of 
equation (\ref{outside}). This transverse term causes  
 further decrease of and subsequent exponential development 
of the Lorentz factor. Thus, we find  that the ideal-MHD condition does not
 hold between the inner boundary and the stellar surface. 
On these ground, we conclude that other electrons populate   
between the stellar surface and the inner boundary to hold the condition  
$E_{||}=0$ (i.e. the ideal-MHD condition). These electrons will be 
supplied from the stellar surface or the electronic cloud on the polar caps  
by the negative field-aligned electric field, which will appear unless there 
are such electrons.

The same argument is applied for  outside of the outer boundary, and 
therefore the ions (or positrons) in addition to the positrons 
escaping from the gap are required.
\subsection{Displacement of the outer boundary}
\label{displace}
In all the four cases in \S\ref{result}, we assume that each position of the 
outer boundary is located far outside from the null surface. 
Then expansion of the gap in the trans-field direction is suppressed 
by the pair creation. As a result,  the trans-field thickness $D_{\perp}$ 
becomes shorter than the gap width $W_{||}$.

In our model, the location of the outer boundary $\zeta_{out}$ is 
free (\S\S\ref{method}), but determine the value of the total current running 
through the gap. We may put the outer boundary near
the null surface.  In such  cases, because $W_{||}$ is short,
we have a gap which has a larger $D_{\perp}$  than
those of Case~1$\sim$Case~4.  In this \textit{wedge}-like gap,
because the trans-field propagating distance of the $\gamma$-rays in the gap
is negligible as compared with $D_{\perp}$, the magnetic field lines are 
effectively straight in the gap. We then anticipate that the solution 
of our two-dimensional model with this geometry 
is effectively the same as  that of the one-dimensional model.

\begin{figure}
\begin{center}
\includegraphics[width=10cm, height=8cm]{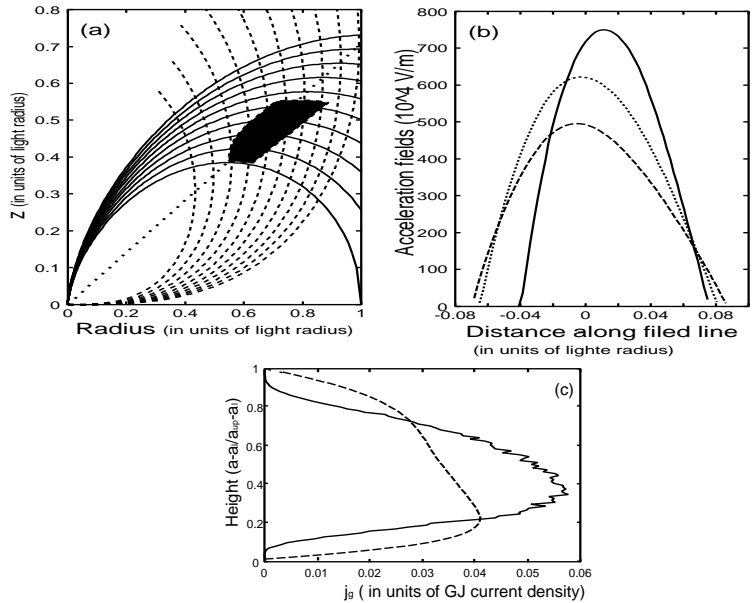}
\caption{A solution of the wedge-like gap. The external currents $(j_1,j_2)=
(0.1,0.2)$ is the same as Case~2. (a) The geometry of the
outer gap (dark region). The solid, dashed, and dotted lines are the same as
 in Fig.\ref{nvagap}. (b) The distribution of the field-aligned electric field.
 The solid, dotted, and dashed lines are the same as in Fig.\ref{electric}. (c)
The trans field structures of the current and the field-aligned electric field.
 The solid and dashed lines are the same as in Fig.\ref{current}.}
\label{wedgap}
\end{center}
\end{figure}

Fig.\ref{wedgap} shows a solution of the wedge-like gap, where
 the model parameters $j_1=j_2=0.1$, which are  the same as Case~2. As 
a result, there is no solution if $j_g$ becomes greater than 
$0.1$, which is smaller than the critical value of $\sim 0.17$ for  Case~2.  
Furthermore, the distribution of the field-aligned electric field is 
quadratically, and is difference from one of the slab-like gap such 
as Case~1$\sim$Case~4.
\subsection{Drift motion}
\begin{figure}
\begin{center}
\includegraphics[width=6cm, height=6cm]{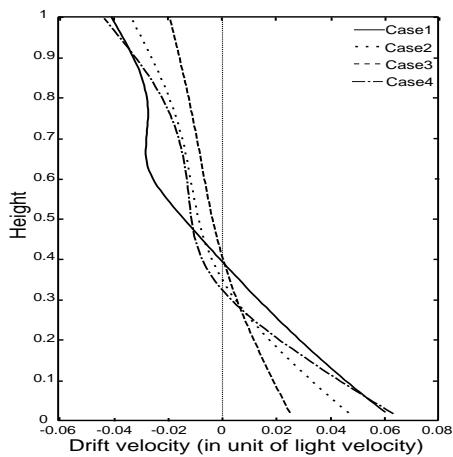}
\caption{The non-corotational drift motion around the rotation axis on the 
outer boundaries. The solid, dotted, dashed, dashed-dotted correspond to 
the Case~1, 2, 3, and 4, respectively. The abscissa is the drift velocity 
in units of light velocity. }
\label{drift}
\end{center}
\end{figure}

Let us examine validity of the assumption 
that the non-corotational drift motion is negligible as compared with 
the corotational motion. From equation (\ref{drifteq}), the drift motion 
around the rotational axis may be expressed as 
\begin{equation}
v_{\phi}=\varpi\Omega+c\frac{E_{\perp}}{B},
\label{drifteq1}
\end{equation}
 where $E_{\perp}=-\partial \Phi_{nco}/\partial s_{\perp}$. The first and 
the second terms of the right hand side of the equation (\ref{drifteq1}) 
represent the corotational and the non-corotational motion, respectively.
Fig.\ref{drift} shows the drift velocity of the  non-corotational part 
evaluated on the outer boundary. 
The plus (or minus) sign for the abscissa represents 
the direction (or counter-direction) of the stellar rotation. From 
this figure, we find that the drift motion becomes super-corotation 
in the lower part of the gap, and sub-corotation  in the upper part.
In our calculation, we find that non-corotational motion is a few per 
cent of the corotational motion (i.e. $|\mathbf{E}_{co}|\gg|E_{\perp}|$).
The rotational period $\hat{P}_3$ of drift motion of plasmas
 due to  $E_{||}$ in the outer gap is suggested to be
$\hat{P}_3\sim13P_1$, where $P_1$ is the corotational period, and
we used the maximum drift velocity $|v|\sim0.06c$ at $\varpi=0.8\varpi_{lc}$. 
Since we have neglected the structure of the azimuthal direction, we cannot 
calculate the drift velocity due to the azimuthal electric field. 
However, we may anticipate that the strength of the azimuthal electric field 
are same order (or less) as compared with  the values of the trans-field 
electric field $E_{\perp}$, unless the dimension in the 
azimuthal direction is much smaller than that in the meridional plane.
  Therefore, we safely neglect the non-corotational drift motion.

In the polar cap model of Ruderman \& Sutherland (1975), 
the subpulse drift of the radio band is attributed to $cE_{\perp}/B$ 
drift in the polar cap region. The sub-pulse drifts have been observed 
from the old pulsars such as PSR~B0943+10 ( $\hat{P}_3/P_1\sim37.35$),
 B0826-34 ( $\sim14$), B2303+30 ($\sim23$), 
B0031-07  ($>34$) (Gil, Melikidze \& Geppert 2003). 
The pulsars quoted  are old ($\tau\sim 10^7$yr) and slow 
($P\sim 1$s) pulsars with spin-down energy many orders of magnitude less 
than  known $\gamma$-ray pulsars ($\tau\sim 10^3\sim10^5$yr). 
Even though the pulsar is old, the magnetic 
pair creation process in polar cap region 
must take place due to the high magnetic field ($B_s\sim 10^{12}$G).
 In the outer gap of such old pulsars, 
it is not clear whether the high energy radiation and the photon-photon 
pair creation process take place efficiently.  
In following, therefore, we investigate the outer gap of the old pulsars.  

For simplicity, we assume that the available potential drop, 
$\phi_{a}=B_s\Omega^2R_*^3/2c^2$, on the stellar surface is 
used only for the acceleration of particles in the outer gap. This assumption 
is not valid for the young fast pulsars, 
but not bad for the old and slow pulsars because the gap covers 
the almost pulsar magnetosphere.  
The corresponding  Lorentz factor will be estimated from 
 $m_ec^2\Gamma_a=e\phi_a$, to be $\Gamma_a\sim 2.5\times10^7 P^{-1.5}
\dot{P}^{0.5}_{-15}$, where $P$ is the stellar rotation period in units 
of 1s, and $\dot{P}_{-15}$ is the period derivative in units of $10^{-15}$
$\mathrm{s\cdot s^{-1}}$.
  The pair creation condition is $E_{\gamma}E_{X}\sim2(m_ec^2)^2$, 
where $E_{X}$ is the energy of the soft photons, and $E_{\gamma}$ 
is the curvature photon energy, which is 
$E_{\gamma}\sim 0.1P^{-5.5}\dot{P}_{-15}^{1.5}$GeV, where we adopt 
the light radius as the curvature radius of the magnetic field lines. 
From this condition, the threshold energy of soft photons in terms 
of $P$ and $\dot{P}_{-15}$ becomes $E_X\sim5P^{5.5}\dot{P}_{-15}^{-1.5}$keV. 
For any neutron star cooling models and the polar-cap heating models, 
such high thermal temperature for old slow 
 ($P\sim 1$s, $\dot{P}_{-15}\sim1$) pulsars is not 
predicted (Zhang et al. 2004, and references therein).  For checking, we 
applied the present numerical code for an older pulsar PSR 0943+10 with 
the X-rays field of an unexpectedly temperature, which we adopted for Vela pulsar ($E_X\sim0.15$kev).  We found that  no pairs are
 produced in the gap. On these ground,  we conclude that the 
pair creation process does not take place in the outer gap of 
the old pulsars quoted above. 

Such silent gap does not produce current carrying particles by itself. 
However if the external current comes into the gap through the boundaries, 
the electrons (or positrons)   are accelerated in the gap, 
and then radiate curvature photons (for example, the energy of the 
$\gamma$-rays is about $E_{\gamma}\sim 100$MeV with $P\sim 1$s and 
$\dot{P}_{-15}\sim 1$).  If the external current has   
the Goldreich-Julian value $B_s\Omega^2R_{*}^3/c$, then 
the  luminosity is same order of the spin down energy, namely 
$L_{\gamma}\sim I_{GJ}\times\phi_a\sim B_0^2\Omega^4R_*^6/c^3$. For example, 
the predicted luminosity for PSR B0826-34, the observed distance 
($d\sim0.54$kpc, Lyne \& Smith 1998) of which is the nearest in the pulsars 
quoted above, is $\sim 6.2\cdot 10^{30}$ erg/s.  However, 
since the corresponding flux on Earth,  
$F\sim L_{\gamma}/4\pi d^2$, is $\sim 2\cdot 10^{-13}$ erg/$\textrm{cm}^{2}$s, 
which is below  sensitively of and unobservable even with the 
new $\gamma$-ray telescope GLAST. 
  On the other hand, the middle age pulsars ($\tau\sim 10^6$yr) 
will have the gap in which  both of the pair creation and the 
$\gamma$-ray radiation processes take place, and 
are interesting candidates for GLAST (Torres \& Nuza 2003).

\section*{Acknowledgments}
We are grateful to Dr Motokazu Takizawa for fruitful discussions. 
Authors also thank Dr Geoffrey Wright, the referee, for insightful 
comments on the manuscript.

\label{lastpage}
\end{document}